\documentclass[twocolumn,pre,aps,showpacs,epsfig,dvips,amsmath,amssymb]{revtex4}
\usepackage{eufrak}
\usepackage[dvips]{graphicx}
\usepackage{dcolumn}
\usepackage{bm}
\usepackage{multirow}

\newcommand{\nh}{{\hat{n}}}

\begin{document}
\title{Topology of Smectic Order on Compact Substrates}
\author{Xiangjun Xing}
\affiliation{Department of Physics, Syracuse University, Syracuse, New York~13244}
\date{\today} 

\begin{abstract}
Smectic orders on curved substrates can be described by differential forms of rank one (1-forms), whose geometric meaning is the differential of the local phase field of density modulation.  The exterior derivative of 1-form is the local dislocation density.   Elastic deformations are described by superposition of exact differential forms.   Applying this formalism to study smectic order on torus as well as on sphere, we find that both systems exhibit many topologically distinct low energy states, that can be characterized by two integer topological charges.  The total number of low energy states scales as the square root of the substrate area.   For smectic on a sphere, we also explore the motion of disclinations as possible low energy excitations, as well as its topological implications.  

\end{abstract}

\maketitle

\noindent

The problem of smectic order on curved substrate naturally arises in a variety of contexts, such as block copolymer films coated on a curved substrate~\cite{santangelo:017801,chantawansri:lamellar-sphere,copolymer-sphere-simulation-Li,Tang-2006}, colloidal particles immersed in a smectic liquid crystal~\cite{Kleman-smectic-confinement-01} 
with strong tangential boundary condition, smectic polymer vesicles, as well as Turing patterns on a sphere~\cite{Turing-pattern-ref}.    Last but not least, a flexible charged polymer adsorbed onto an oppositely charged curved surface~\cite{Charged-polymer-ref} may also form equal-distanced layer pattern on the surface. 

The main purpose of this letter is to study the low energy smectic states on curved substrates {\em with minimal number of defects}, which are experimentally most relevant at low temperature.   We shall find that smectic order both on sphere and on torus exhibit many topologically distinct, nearly degenerate, and minimally defected states.  These states can be classified by two integer topological charges, which are related to substrate topology as well as to disclination pattern.   The total number of these states is proportional to $\sqrt{A}/\ell_0$, where $A$ is the substrate area, $\ell_0$, the preferred layer spacing.   Their energy differences scale sub-linearly with the system size.  
For smectic on sphere, by exchanging defects pairs in an appropriate manner, two integer charges can be systematically changed.  Interestingly, the corresponding process realizes the Euclidean algorithm for finding the greatest common divisor of two integers.  Our results are complementary to a recent work by Santangelo {\it et al}~\cite{santangelo:017801}, which addresses the energetic interplay between smectic order and substrate curvature.   A more detailed version of our analysis is presented elsewhere~\cite{Xing-curved-smectic-long}.


The local phases of density modulation of a smectic order at two nearby points $x$  and $x + d x $ are related by  
$\Theta(x+  \delta x) \approx  \Theta(x) + \psi_{\alpha}(x) d x^{\alpha}.$
While the phase field $\Theta$ is generically not globally well defined, its differential 
$\psi(x) = \psi_{\alpha} dx^{\alpha}$, called a 1-form in modern differential geometry~\cite{footnote-exterior}, nevertheless still is.  A simple geometric reasoning shows that the norm of $\psi$ is proportional to the reciprocal of layer spacing $\ell$: 
$|\psi| = \sqrt{ g^{\alpha\beta} \psi_{\alpha} \psi_{\beta}} 
= 2\pi\,\ell^{-1}.  $
The nonlinear strain
$w(x) = \frac{1}{2} \left(\frac{\ell_0^2}{4\pi^2} |\psi|^2 - 1 \right)$
measures the dilation/compression of smectic layers.  This strain is identical to the one used in reference \cite{santangelo:017801} and reduces to the well known nonlinear strain for smectic order in flat space~\cite{LC:deGennes,CMP:CL}.  The total free energy of the system is the sum of the strain energy density $B\, w^2 /2$ and the Frank free energy, appropriately generalized to curved space~\cite{footnote-detail}. 

The integral of $\psi$ along a loop $\gamma$ is the net phase change $\Delta \Theta$ as one walks around $\gamma$, i.e. the dislocation charges, multiplied by $2 \pi$, enclosed by the loop.   Using the Stoke's theorem this can in turn be related to the 2d integral over the region $D$ enclosed by $\gamma$: 
\begin{eqnarray} 
\Delta \Theta = 2 \pi N_{\gamma} =  \oint_{\gamma} \psi =  \int_{D} d\psi ,
\label{Delta-Theta}
\end{eqnarray}
where the two form $d\psi =  (\partial_x \psi_y - \partial_y \psi_x)dx \wedge dy$ is the exterior differential of $\psi$. $d \psi$ therefore must be the dislocation density.    Hence a dislocation-free smectic state is described by a 1-form satisfying $d \psi = 0$.  Such a 1-form is called {\em closed}.   By  contrast, a 1-form $\psi = d \Theta$ that is the differential of a function $\Theta$ is called {\em exact}. 

\begin{figure}
\begin{center}
\includegraphics[width=8.5cm]{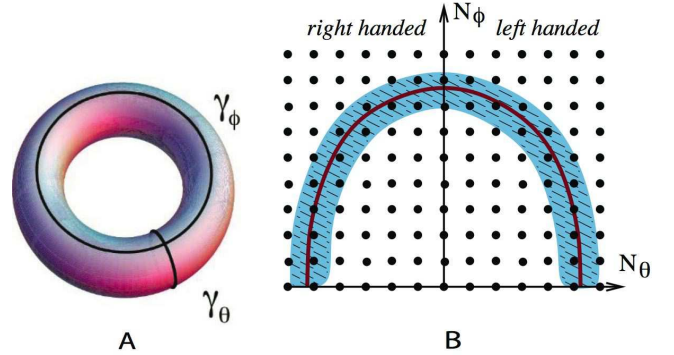}
\caption{ A): Two cycles on a torus.  B)
Smectic states on a torus with distinct global dislocation charges are shown as grid points in the upper-half $(N_{\theta},N_{\phi})$ plane.  The curve (ellipse) is the loci of vanishing strain $w = 0$, as determined by Eq.~(\ref{vanish-strain-locus}).  States in the shaded region have low strain energy and are approximately degenerate. }
\label{torus-cycles}
\end{center}
\vspace{-5mm}
\end{figure}

An exact form is closed.  If the phase field $\Theta$ is globally defined, we have $\psi = d \Theta$, and $d \psi = d^2 \Theta  = (\partial_1 \partial_2 - \partial_2 \partial_1) \Theta \, dx^1 \wedge dx^2 \equiv 0$, that is, the dislocation density vanishes everywhere.   This obvious result is called Poincar\'e's lemma.   Its converse holds in flat space: if $d\psi = 0$ everywhere, the phase field $\Theta $ is globally defined: $\psi = d\Theta$, and the integral Eq.~(\ref{Delta-Theta}) along arbitrary loop always vanishes.  

The converse of Poincar\'e's lemma may fail in space with different topology \cite{footnote-failure}.  De Rham's theorem of cohomology \cite{books:math} identifies the number of independent closed-but-not-exact forms with that of non-retractable loops on the manifold.  A torus, for example, has two non-retractable loops, as shown in Fig.~\ref{torus-cycles}A. The most general closed 1-form on torus is given by 
\begin{eqnarray}
\psi_{N_{\theta}, N_{\phi}} = 
N_{\theta} \, d\theta + N_{\phi} \,d\phi
+ d\Phi(\theta,\phi).   
\label{packing-torus} 
\end{eqnarray} 
The integral of $\psi$ along loops $\gamma_{\theta}$ and $\gamma_{\phi}$ give respectively $2\pi N_{\theta}$ and $2 \pi N_{\phi}$.  Eq.~(\ref{packing-torus}) is exact only if both  $N_{\theta}$ and $N_{\phi}$ vanish.  In order for Eq.~(\ref{packing-torus}) to describe a defects-free smectic state on torus, both $N_{\theta}$ and $N_{\phi}$ must be integers.  The exact form $d\Phi(\theta,\phi)$ in Eq.~(\ref{packing-torus}) describes an elastic deformation of smectic layers relative to the state with $d \Phi = 0$.  It can be ignored for a sufficiently thin torus. 

We define two states to be topologically equivalent if they can be brought into each other by continuous deformation that do not break any smectic layer.   The topological properties of a defects free smectic state on a curved substrate are therefore characterized by a closed-but-not-exact differential form.  Similarly, any excitation/distortion of the smectic pattern that can not be relaxed away by elastic deformation is a topological defects.  According to this definition, therefore topological defects should be identified as the source of {\em internal strain}, a concept thoroughly explored in the continuous theory of defects in crystals~\cite{Kroner-notes}.   We note, however, this definition of topological equivalence is more refined than the one used in the homotopy theory topological defects \cite{Kleman-review,Mermin-review}.  Consequently, the topological charges discussed in this work may not be truly topological invariants according to the homotopy theory.   A more detailed discussion of this issue can be found in \cite{Xing-curved-smectic-long}.

Toroidal smectic states with different charges $(N_{\theta}, N_{\phi})$ are clearly topologically distinct. 
Two integers $(N_{\theta},N_{\phi})$ share essential similarity with usual dislocation charges, but encode global properties of {\em defects-free} smectic states on a torus.  Hence they shall be called {\em global dislocation charges}~\cite{footnote-crystal}.   
Each defects-free smectic state can therefore be represented as a point in the half lattice of integers $(N_{\theta},N_{\phi})$~\cite{footnote-inversion}, as shown in Fig.~\ref{torus-cycles}B.

The nonlinear strain associated with states Eq.~(\ref{packing-torus}) on a thin torus (with $d \Phi$ set to zero) is given by:
\begin{eqnarray}
w =  \frac{\ell_0^2 N_{\theta} ^2 }{(2 \pi R_{\theta})^2} 
+ \frac{\ell_0^2 N_{\phi} ^2 }
{(2 \pi R_{\phi})^2} - 1. 
\label{vanish-strain-locus}
\end{eqnarray} 
As shown in Fig.~\ref{torus-cycles}B, the equation $w = 0$ traces out an ellipse in the $(N_{\theta},N_{\phi})$ plane.   States satisfying $|w| \leq \ell_0/\sqrt{A}$ (shaded in Fig.~\ref{torus-cycles}B) have the total strain energy bounded by $B\,\ell_0^2/2$, which is independent of the system size.  A simple calculation~\cite{Xing-curved-smectic-long} also shows that the total Frank free energy scales sub-linearly with the system size.  
Therefore all these states are approximately degenerate for a large system.  The total number of these low energy states is given by the area of the shaded region in Fig.~\ref{torus-cycles}B, and scales as the square root of the substrate area: 
\begin{eqnarray}
{\mathcal N}(F_B \leq  \frac{1}{2} B\, \ell_0^2) = 2 \pi \frac{\sqrt{A}}{\ell_0}, 
\end{eqnarray} 
 
 
We now turn to smectic order on sphere.   According to the Gauss-Bonnet-Poincar\'e theorem, the total disclination charge of a smectic order on sphere must be two.  Let us write $\psi = |\psi| \nh $, where $\nh = \nh_{\alpha} dx^{\alpha}$ is the unit 1-form describing the smectic layer normal, while $|\psi|$ is the norm of $\psi$.  Taking the exterior differential of $\psi$, as well as using the Leibniz rule, we find 
$d \psi =  (d|\psi|) \wedge \nh + |\psi| d \nh.$
Note that $d\nh =  (\partial_1 \nh_2 - \partial_2 \nh_1) dx^1 \wedge dx^2 $ is a 2-form describing bending deformation of the layer normal.  Hence if there is no dislocation, $d\psi  = 0$, and the layer spacing is constant, $d|\psi| = 0$, the bending deformation of the director field is strictly forbidden, $d\nh = 0$.  This implies that the bending constant $K_3$ is effectively infinity in a dislocation-free smectic \cite{footnote-bending}.   

A spherical nematic in the limit of infinite bending rigidity  
is characterized by a one-parameter family of degenerate ground states, where four $+1/2$ disclinations sit on a great circle and form a rectangle with arbitrary aspect ratio~\cite{SBX-sphereical-nematics}.  For a given bending-free nematic state, we can start from disclination cores and grow, layer by layer, a dislocation-free smectic pattern with equal layer spacing.  We will, however, have to fine tune the layer spacing $\ell$ so that the smectic pattern can be fit onto the sphere with given radius $R$.  This fine tuning results in a small strain of order of $\ell_0/R$, and a total strain energy  $F_B \approx   B \, \ell_0^2$,  which does not scale with the system size.   Since the Frank free energy of all these states is the same by construction, the total free energy is approximately degenerate.   

\begin{figure}
\begin{center}
\includegraphics[width=8cm]{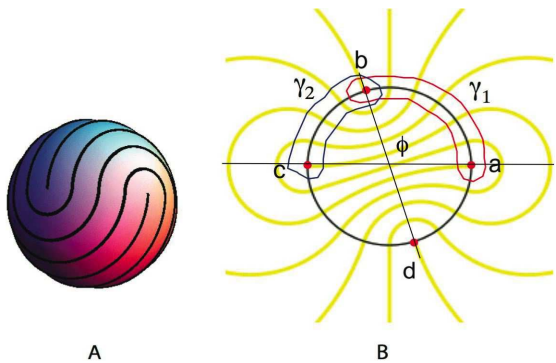}
\caption{ (Color online) (A) A quasi-baseball state, with four $+1/2$ disclinations sitting on the equator.  Each disclination core has one layer termination.  Only two disclinations are visible here.  
(B) The same state stereographically projected onto the complex plane.  The equator is mapped to the unit circle.   The red contour encloses defects $a, b$, and intersects $12$ smectic layers in total. The blue contour encloses defects $b, c$ and intersects $8$ layers in total.  Hence this quasi-baseball state is labeled by two integer charges $(N_1=12, N_2 = 8 )$.  }
\label{branchcuts-abcd}
\end{center}
\vspace{-5mm}
\end{figure}

Several numerical studies of spherical smectic orders formed by di-block copolymer films confined on sphere have been reported recently~\cite{chantawansri:lamellar-sphere,copolymer-sphere-simulation-Li,Tang-2006}.   By slowly annealing the system, three classes of low energy smectic states were found~\cite{chantawansri:lamellar-sphere}: latitudinal states~\cite{footnote-Latitude}, where all layers are circles of constant latitude,  spiral states,  where all smectic layers are spirals around the two poles, and quasi-baseball states, where four $+1/2$ disclinations are well separated on a great circle.   All these states are dislocation-free.  There may be, however, at most one layer termination at each disclination core.

Consider a quasi-baseball state, shown in Fig.~\ref{branchcuts-abcd}A, with four disclinations sitting on the equator.  We stereographically project the pattern from the south pole to the complex plane, such that the equator is mapped into the unit circle.  The image smectic pattern is shown in Fig.~\ref{branchcuts-abcd}B, with four defects located at 
$(a,b,c,d) = (1, e^{i \phi}, -1, e^{-i \phi})$ respectively. 
The segment $ab$ of the unit circle intersects $N_1/2$ smectic layers.  (A layer terminated at disclination $a$ or $b$ is counted as $1/2$ intersection. )  Likewise, the segment $bc$ of the unit circle intersects $N_2/2$ smectic layers.  Note that by definition both $N_1$ and $N_2$ are non-negative.  Without loss of generality we choose $N_1 \leq N_2$.   Two integer charges $(N_1,N_2)$ completely determine the global properties of a quasi-baseball state.  Latitudinal states are special cases of quasi-baseball states with $N_1 = 0$~\cite{footnote-N10}, while spiral states correspond to the regime $N_1 \ll N_2$.

It is clear from Fig.~\ref{branchcuts-abcd}A that all smectic layers intersect the equator vertically.  The arc lengths $s_{ab}$ and $s_{bc}$ are therefore given by  
\begin{eqnarray}
s_{ab} = R \, \phi = N_1 \ell,  \quad
s_{bc} = R \, (\pi - \phi) = N_2 \ell, \nonumber
\end{eqnarray}
from which we find 
\begin{eqnarray}
\frac{N_1}{N_2} = \frac{\phi}{\pi - \phi}, 
\quad
\ell = \frac{\pi R}{N_1 + N_2},
\label{d-R-N}
\end{eqnarray}
where $\ell \approx \ell_0$ is the layer spacing.  The equilibrium value of the angle $\phi$ is therefore completely determined by the two charges $N_1$ and $N_2$.  Large fluctuations of $\phi$ necessarily induces significant change of layer spacing, and therefore is energetically penalized.    More importantly, Eq.~(\ref{d-R-N}) also shows that, in all low energy states, $N_1+ N_2$ is completely determined by the sphere radius $R$ and the layer spacing $\ell$.  Since $N_1$ can take arbitrary integer value from $0$ to $ \pi R /\ell$, the total number of low energy smectic states on sphere is approximately given by 
\begin{eqnarray}
{\mathcal N}(F_B \leq  B\, \ell_0^2) 
= \frac{\pi R}{\ell_0} + 1 
\approx \frac{\sqrt{\pi A}}{\ell_0},
\end{eqnarray}
i.e. also scales as the square root of the sphere area.

To obtain an analytic description for quasi-baseball states, we invoke the Hodge decomposition theorem~\cite{books:math} of differential forms on compact manifold.  The closed 1-form $\psi$ can be expressed as the sum of the real part of an analytic form $\psi^c$ and an exact form $d\alpha$:
\begin{eqnarray}
\psi(x, y) = Re \, \psi^c(z) + d \alpha(x,y), 
\end{eqnarray}
where $\psi^c(z) = f(z) dz$, with $f(z) = f(x+i y)$ a meromorphic function, while $\alpha(x,y)$ is a smooth (but not analytic) function on the z-plane.  The topological properties of the smectic order on sphere is completely encoded in $f(z) dz$.   The exact form $d\alpha$ describes elastic deformation and should be chosen to minimize the total free energy.   

The complex meromorphic form $\psi^c$ that describes a spherical smectic with four disclinations at $a,b,c,d$ can be shown to be~\cite{Xing-curved-smectic-long} 
\begin{eqnarray}
\psi^c = \frac{A e^{-i\,\alpha}}{\sqrt{(z-a)(z-b)(z-c)(z-d)}} \, dz
\label{psi-abcd}
\end{eqnarray}
As shown in Fig.~\ref{branchcuts-abcd}B, let us construct two loops $\gamma_1$ and $\gamma_2$ that enclose the segments $ab$ and $bc$ respectively.  Integration of $\psi$ along the loops $\gamma_1,\gamma_2$ yield two topological charge $N_1,N_2$ that we defined above:   
\begin{eqnarray}
2 \pi \, N_k =  \oint_{\gamma_k} \psi 
= Re \, \oint_{\gamma_k} \psi^c, 
\quad k = 1,2. 
\label{na-sphere}
\end{eqnarray} 
Solving Eq.~(\ref{na-sphere}) for $A$ and $\alpha$ we obtain
\begin{eqnarray}
A \, e^{- i \alpha} = \left( \frac{\pi N_2}{K\left( \cos \frac{\phi}{2} \right)} 
 - i \frac{\pi N_1}{K\left( \sin \frac{\phi}{2} \right) } \right)\, e^{i \phi/2}, 
 \label{A-solutions}
\end{eqnarray}
where $K(\cdot)$ is complete elliptic integral of the first kind.   Eqs.~(\ref{psi-abcd}) and (\ref{A-solutions}) completely determine the meromorphic form $\psi^c$ for given charges $N_1,N_2$.   We note that the integral of $\psi^c$ Eq.~(\ref{psi-abcd}) defines the (multiple valued) inverse function of a doubly periodic elliptic function.  Two integer charges $N_1,N_2$ are the real parts of two periods of this elliptic function~\cite{Xing-curved-smectic-long}.


\begin{figure}
\begin{center}
\includegraphics[width=6.5cm]{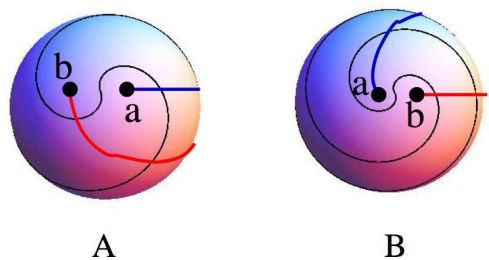}
\caption{(Color online) Starting from a state $(N_1, N_2)$ with $N_1 = 2$, we twist the sphere counter-clockwise by $\pi$ so as to exchange the defects pair $a,b$, and obtain a state $(N_1, N_2+N_1)$.  These two states are therefore topologically equivalent.  This twist operation changes the layer spacing by a factor of $N_1/N_2$.  
}
\label{sphere-twist-transform}
\end{center}
\vspace{-5mm}
\end{figure}

Spiral state are especially interesting because of their closely bound defects pairs.  As illustrated in Fig.~\ref{sphere-twist-transform}, consider a spiral state with charges $(N_1 = 2,N_2)$, 
align the sphere so that one pair of defects $a,b$ sit near the north pole.  Let us fix the south pole and elastically twist the whole sphere, together with all smectic layers, around the z axis by an angle $\pi$,  so that the defects pair $a,b$ at the north pole exchange their positions.  The branch cut (shown blue in Fig.~\ref{sphere-twist-transform}A) which connects $a$ and $d$ (invisible in Fig.~\ref{sphere-twist-transform}) before the twist operation, after the twist operation, connects $b$ and $d$ instead, shown red in Fig.~\ref{sphere-twist-transform}B.  The corresponding integer charge associated with this branch cut is $N_2$ before the twist, and becomes $N_2 + N_1$  after the twist.  Hence a counter-clockwise twist of the defects pair $a,b$ by angle $\pi$ leads to the following transformation of the topological charges 
\begin{eqnarray}
(N_1,N_2) \rightarrow (N_1,N_2 +N_1). 
\label{transform-1}
\end{eqnarray}
Obviously clockwise twist of the pair $(a,b)$ leads to a different transformation:
\begin{eqnarray}
(N_1,N_2) \rightarrow (N_1,N_2 - N_1). 
\label{transform-2}
\end{eqnarray}
Therefore, if defects are allowed to move, states $(N_1,N_2 \pm N_1)$ are topologically identical to the state $(N_1,N_2)$.   As long as $N_1 \ll N_2$, this twist manipulation only induces small change in layer spacing and therefore costs little strain energy.  Twist of closely bound defects pair is therefore a low energy excitation in spiral states.  

It is important to note that even if $N_1$ and $N_2$ are comparable, twist of defects pair is still topologically possible, but costs large strain energy.   The two charges $(N_1,N_2)$ of a quasi-baseball states are therefore stabilized by both topological and energetic barriers.  Furthermore, 
defects twisting can be successively applied.  Starting with an state with $N_1 \leq N_2$, we twist an appropriate defects pair so as to obtain new charges $(N'_1 = N_1,N'_2 = N_2-N_1)$.   If $N_1' \leq N_2'$ still holds, we simply repeat the same twist operation.   If, after the twist, we find $N_1' \geq N_2'$, we twist a different pair so that we have a new state $(N_1', N_2' - N_1')$.   In every step, one of two integers $(N_1,N_2)$ is reduced.  This process therefore must end after finite steps, where one of the two integer charges vanishes.  The final state is therefore a latitudinal state.   More interestingly, this process of charge reduction is precisely the {\em Euclidean algorithm} for finding the greatest common divisor $gcd(N_1,N_2)$ between two integers $(N_1,N_2)$.  Therefore the nonvanishing charge of the final state must be $gcd(N_1,N_2)$.  Hence if disclinations are allowed to move, an arbitrary low energy smectic state on a sphere is topologically identical to certain latitudinal state, with only one nonzero integer topological charge.   


\begin{acknowledgments}
The author acknowledges helpful discussions with Mark Bowick, and graphic assistances from Luca Giomi, as well as financial support from the American Chemical Society under grant PRF 44689-G7.   
\end{acknowledgments}


\begin{thebibliography}{10}
\bibitem{footnote-failure}
Such a failure can only be seen at the global scale.  Locally a manifold resembles flat space, hence the converse of Poincar\'e lemma always holds locally.
\bibitem{footnote-inversion}
The state $(-N_{\theta},-N_{\phi})$ is the same as $(N_{\theta},N_{\phi})$.
\bibitem{footnote-bending}
This result applies to arbitrary surface, since the above discussion does not refer to any metric structure.
\bibitem{footnote-Latitude}
These are called hedgehog states in reference~\cite{chantawansri:lamellar-sphere}.
\bibitem{footnote-N10}
For a latitudinal state, $N_1 = 0$, the contour $\gamma_1$ in Fig.~\ref{branchcuts-abcd} can be shrunk to a single point and gives no information about the global property.  There is then only one contour, $\gamma_2$ left, which gives one integer charge.
\end{thebibliography}
\end{document}